\documentclass{article}
\usepackage{spconf,amsmath,graphicx}
\usepackage{xcolor}
\usepackage{url}
\usepackage{multirow}
\usepackage{amssymb}
\usepackage{pifont}
\usepackage{enumitem}
\usepackage{hyperref}
\usepackage{siunitx}
\usepackage{booktabs}
\usepackage{caption}
\newcommand{\cmark}{\ding{51}}%
\newcommand{\xmark}{\ding{55}}%

\title{Complexity Scaling for Speech Denoising}
\name{Hangting Chen, Jianwei Yu, Chao Weng}
\address{Tencent AI Lab, Audio and Speech Signal Processing Oteam}
\begin{document}
\ninept
\maketitle
\begin{abstract}
Computational complexity is critical when deploying deep learning-based speech denoising models for on-device applications. Most prior research focused on optimizing model architectures to meet specific computational cost constraints, often creating distinct neural network architectures for different complexity limitations. This study conducts complexity scaling for speech denoising tasks, aiming to consolidate models with various complexities into a unified architecture. We present a Multi-Path Transform-based (MPT) architecture to handle both low- and high-complexity scenarios. A series of MPT networks present high performance covering a wide range of computational complexities on the DNS challenge dataset. Moreover, inspired by the scaling experiments in natural language processing, we explore the empirical relationship between model performance and computational cost on the denoising task. As the complexity number of multiply-accumulate operations (MACs) is scaled from 50M/s to 15G/s on MPT networks, we observe a linear increase in the values of PESQ-WB and SI-SNR, proportional to the logarithm of MACs, which might contribute to the understanding and application of complexity scaling in speech denoising tasks. \footnote{Demos and detailed results are on \url{https://hangtingchen.github.io/Complexity-Scaling-for-Speech-Denoising.github.io/}.}

\end{abstract}
\begin{keywords}
Speech denoising, neural network, multi-path transformer, complexity scaling
\end{keywords}
\section{Introduction}
\label{sec:intro}

Speech denoising aims to remove environmental noise to improve speech quality and intelligibility, which is an important task for telecommunications and hearing aids. Over the last decades, deep learning-based models have achieved promising performance on the noise suppression task.

Following the successful Deep Noise Suppression challenges \cite{Dubey2022Icassp2D,Reddy2021Interspeech2D,Reddy2020ICASSP2D}, many classic models have been proposed. These models can be classified into two directions: pursuing high performance with large models and enhancing the model's capability under specific limitations. Some popular backbone architectures, such as DCCRN \cite{DBLP:conf/interspeech/HuLLXZFWZX20}, full-sub-band \cite{DBLP:conf/icassp/HaoSHL21,DBLP:conf/icassp/DangCZ22} and band-split models \cite{DBLP:journals/taslp/LuoY23}, own large computational complexities. Meanwhile, several studies of low-complexity denoising models have been conducted, including DeepFilterNet \cite{DBLP:conf/icassp/SchroterERM22}, Fast FullSubNet \cite{hao2022fast} and ultra dual-path compressed DPT-FSNet \cite{chen23t_interspeech}. 
One limitation of the previous research is that most of them cover a narrow range of computational costs. To satisfy various resource requirements for real-world applications, researchers must design models with different target computational costs, which might lead to distinct neural network architectures. 
Thus, a unified model architecture, evaluated to be effective in a wide range of computational complexities, will largely save the time cost on the architecture design. 

On the other hand, building models on a unified model architecture promotes the discovery of scaling law on front-end signal processing. Neural scaling law indicates empirical relationships between parameters, facilitating the designing and understanding of neural network models. Many research fields have explored the neural scaling law, for example, vision transformer on ImageNet \cite{DBLP:conf/cvpr/Zhai0HB22}, transformer models on neural machine translation \cite{DBLP:conf/iclr/GhorbaniFFBKGCC22} and language models on WebText2 \cite{DBLP:journals/corr/abs-2001-08361}. However, few studies have been conducted to present scaling laws in the field of speech enhancement.

\begin{figure}
\resizebox{0.45\textwidth}{!}{
\centerline{\includegraphics[width=\columnwidth]{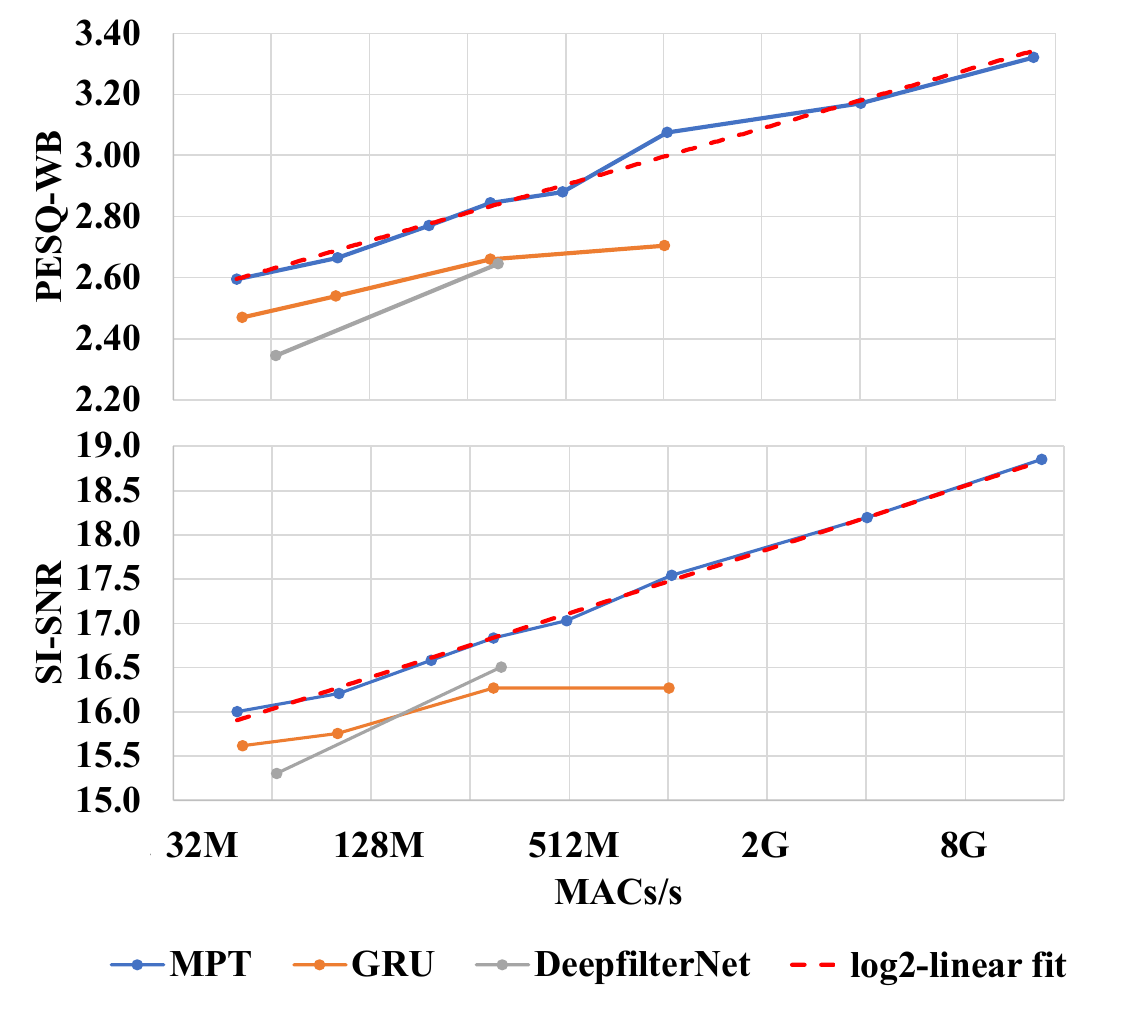}}}
\caption{The objective metrics of PESQ-WB and SI-SNR increase smoothly as we increase the computational cost. The metrics are evaluated on DNS-2020 \textit{no\_reverb} / \textit{with\_reverb} test sets.}
\label{fig:results}
\end{figure}

This study proposes a Multi-Path Transformer (MPT) network to satisfy different computational complexities towards the denoising tasks. Meanwhile, changing the model configuration illustrates a scaling law between computational cost and performance for the first time. Our main contributions are two-fold:
\begin{itemize}[itemsep=0pt,topsep=0pt,parsep=0pt,leftmargin=10pt]
\item 
\textbf{MPT architecture} can handle various complexities by adapting MPT blocks for multi-view modeling and using two streams to generate estimations. When scaling from high to low computational complexity, the MPT network achieves competitive performance compared to previous architectures.
\item 
\textbf{Scaling law} between computational complexity and performance is formulated according to the performance of MPT networks (Figure \ref{fig:results}). The empirical formulas can be described in power law, delivering intuitions on how objective metrics change when varying the computational cost.
\end{itemize}
To the best of our knowledge, the proposed MPT network is the first one to cover multiply-accumulate operations (MACs) from 50M/s to 25G/s and to show competitive performance on all computational complexity when tested on the DNS challenge data. Meanwhile, the performance diagram in Figure \ref{fig:results} illustrates that wideband perceptual evaluation of speech quality (PESQ-WB) and scale-invariant signal-to-noise ratio (SI-SNR) will have around 0.09 and 0.36 dB improvement as the computation cost is doubled when the multiply-accumulate operations per second (MACs/s) are below 15G.


\section{Methods}
\label{sec:method}

\begin{figure}
\centerline{\includegraphics[width=0.7\columnwidth]{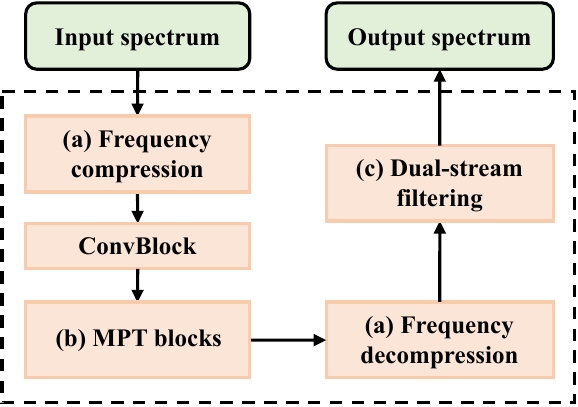}}
\caption{The overview of the MPT architecture. (a) Frequency compression and (a) frequency decompression are plotted together in Figure \ref{fig2}(a). }
\label{fig1}
\end{figure}

\begin{figure*}
\resizebox{1.0\textwidth}{!}{
\centerline{\includegraphics[width=2.0\columnwidth]{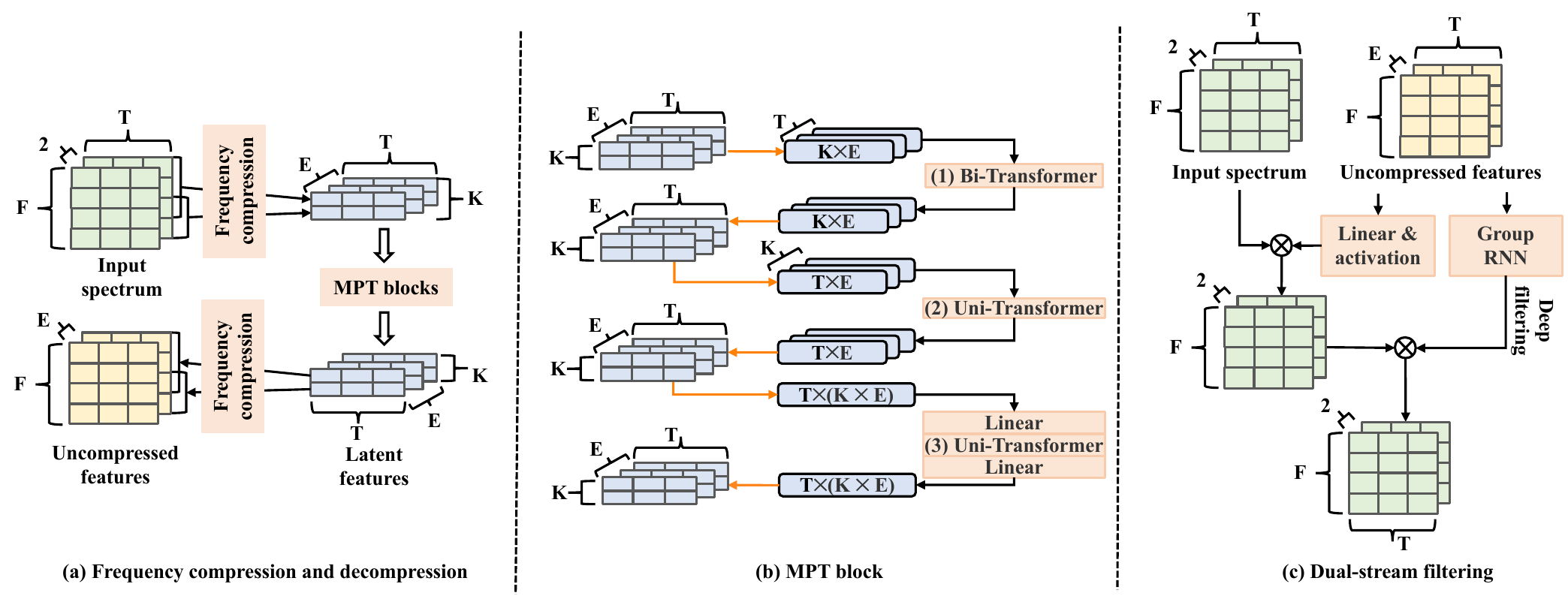}}}
\caption{The detailed diagram of the modules in MPT architecture. The orrange lines in (b) indicate the permutation and reshape operation.}
\label{fig2}
\end{figure*}

In this section, we will first introduce the architecture of the proposed MPT network. Then, we will provide the intuitions behind the settings for different computational complexity models. Some discussions about architecture and neural scaling are also included.

\subsection{MPT network}

Figure \ref{fig1} gives an overview of the model architecture. The frequency compression module transforms the input spectrum into latent features. Then, it goes through a convolution block. The latent features are later passed into MPT blocks. The frequency decompression module decompresses the latent features back into the STFT domain. Finally, a dual-stream filtering module generates the estimated STFT spectrum. A detailed description of each module is provided following the data flow:

\textbf{Frequency compression and decompression} is originated from \cite{chen23t_interspeech} (Figure \ref{fig2}(a)). We use the Mel scale to reduce the number of frequency bands. The frequency compression module transforms the input raw spectra with size $2\times T\times F$ or log power spectra (LPS) with size $1\times T\times F$ into $E\times T\times K$, where 2 represents the stacked of the real and imaginary of the complex spectrum, $T$ is the number of frames, $F$ is the number of frequencies, $E$ is the feature dimension, $K$ is the number of Mel bands. Note that we use LPS as input for low computational cost and stacked raw spectra for high cost, which will be further compared in Section \ref{sec:results}.

For example, with a 16kHz sample rate and 320 fast Fourier transform (FFT) size, 161 frequency bands are compressed into around 30 Mel bands, corresponding to approximately 5x compression. We choose the number of Mel bands around 30 for two reasons. First, our previous experiments \cite{chen23t_interspeech} indicate that when the compression ratio is around 4, the frequency compression achieves similar performance compared with the dual-path compression. Second, works of band-split recurrent neural networks (RNN) also evaluate the efficiency of using such band numbers for a 16kHz sample rate \cite{yu23b_interspeech}. Thus, in this study, we fix the band number around 30. The combination of time compression will be left for future work.

\textbf{ConvBlock} is composed of a causal depthwise separable convolution \cite{DBLP:conf/cvpr/Chollet17} with kernel size $3\times3$ and a 2-dimensional (2D) batch normalization. Different from the pure RNN-based model \cite{yu23b_interspeech}, we remain a convolution block since it provides performance improvement in our experiments. ConvBlock accepts and generates features with size $E\times T\times K$.

\textbf{MPT block} occupies most of the computational cost in MPT networks. Figure \ref{fig2}(b) shows that each MPT block consists of 3 transformers to capture full-band time-invariant, sub-band time-variant and full-band time-variant information. The first one processes each frame with a bi-directional transformer (Bi-Transformer), or called a noncausal transformer, along the band dimension to capture the time-specific full-band information. The second one processes each band with a uni-directional transformer (Uni-Transformer), or called a causal transformer, along the time dimension to capture the sub-band time-variant information. The third one processes the full-band feature with a Uni-Transformer along the time dimension to capture the full-band time-variant information. Additional linear transforms are placed before and after the transformer to project the input with size $K\times E$ into $E$ and subsequently revert it back to the original size. The order of transformers is fixed since changing the order affects the performance slightly.

Uni- and Bi-Transformer have the same architecture with \cite{chen23t_interspeech}, that is, compared with the original version \cite{DBLP:journals/corr/abs-2308-07661}, we use linear attention for fast calculation \cite{DBLP:conf/icml/KatharopoulosV020}, a combination of a recurrent neural network (RNN) and a linear transform as the feed-forward layer for sequential modeling. The RNN in the feed-forward layer \cite{DBLP:journals/corr/abs-2308-07661, DBLP:conf/icml/KatharopoulosV020} will project the input feature with dimension $E$ into a larger size $C\times E$ and then use a linear layer to transform the feature size back. Here, we use a coefficient $C$ to indicate the expansion times. Note that Uni-Transformer uses causal attention and uni-directional RNN while Bi-Transformer uses noncausal attention and bi-directional RNN. MPT blocks accept and generate features with size $E\times T\times K$.

\textbf{Dual-stream filtering} accepts uncompressed features and generates the final estimation. The input uncompressed features with size $E\times T\times K$ are first split into two parts. The first part corresponds to a complex mask or a real-valued idea ratio mask, depending on the input of the compression module, \textit{i.e.,} raw spectra or LPS. The generated mask is applied to the input spectrum to generate the first estimated spectrum. The second part is sent to group RNN to yield deep-filtering coefficients \cite{DBLP:conf/icassp/SchroterERM22, DBLP:conf/iwaenc/SchroterMER22}. The deep filtering is conducted on the first estimated spectrum. We use deep filtering to obtain the final estimation because filtering around the T-F bins will stabilize the prediction and remove undesired interfering counterparts.

The aforementioned MPT is a causal network, which means that each module is causal without using any future STFT frames. We also test a noncausal network with a high computational cost in the experiments, whose RNN and attention layers all adopt bi-directional and noncausal.

\subsection{Scaling model complexity}

The MPT architecture shown in Figure \ref{fig1} has the following hyperparameters: the number of Mel bands $K$, the latent feature size $E$, the number of MPT block $B$, and the projection expansion time $C$. $K$ is set to be around 30 as stated in Section 2.1. We mainly tune hyperparameters $E$, $B$, and $C$ to scale the model computational cost to low and high.

When lowering the computational cost, we conduct two changes to the architecture to alleviate the performance degradation. First, the input complex spectrum introduces larger difficulties compared with magnitude input (i.e., LPS) since the network has to process both the magnitude and phase simultaneously. For low-complexity networks, we use LPS as the input of the frequency compression and estimate the ideal ratio mask. The deep filtering is still conducted on the complex spectrum with complex coefficients. Second, the latent feature size $E$ indicates the modeling redundancy for each band, but the ultra-low computational cost leads to a small $E$. To alleviate this problem, we set the projection expansion time $C$ equal to $1$ and remove the first transform in the MPT block (Figure \ref{fig2}(a)).

\subsection{Discussions on neural architecture and neural scaling}
 \subsubsection{Neural architecture comparison}
Our model inherits lots of strengths from previous works, for example, deep filtering \cite{DBLP:conf/icassp/SchroterERM22, DBLP:conf/iwaenc/SchroterMER22}, transformer block \cite{DBLP:conf/icassp/DangCZ22} and band-split settings \cite{chen23t_interspeech,yu23b_interspeech}. The major differences lie in the MPT block and the dual-stream filtering module, which will be further evaluated in Section \ref{sec:results}. For example, a successful low-complexity neural network is DeepfilterNet, which uses U-net architecture combined with deep filtering. Compared with DeepfilterNet, most layers of MPT are RNNs and linear attention instead of convolutions. Second, our deep filtering is much simpler which does not use the original input spectra.

Another branch is band split and full-sub-band models. For example, band-split RNN (BSRNN) mainly uses RNNs without any convolution \cite{DBLP:journals/taslp/LuoY23}. Compared with BSRNN, the MPT uses attention layers and deep filtering, which will be shown to be effective for low-complexity modeling. Compared with the ultra-compressed online DPT-FSNet \cite{chen23t_interspeech}, we add a third transformer onto the original dual-path module to form the MPT module. Also, we do not include time compression here.

\subsubsection{Limitations in neural scaling exploration}
Neural scaling laws usually explain relationships among losses, network parameters, dataset sizes, and so on \cite{DBLP:journals/corr/abs-2001-08361}. In this study, we explore the scaling laws between network performance and computational costs with a fixed dataset. We study metrics instead of losses since the loss types for training denoising vary in different training frameworks while the objective metrics are usually fixed. We study network complexity instead of size since the computational cost plays a more important role in performance and is also more concerning for on-device applications. We use the fixed DNS dataset, which contains the largest number of clean and noise audios compared to other publicly available datasets.

\section{Experimental setup}
\label{sec:Exp}


\subsection{Dataset}
The experiments were conducted on a simulated dataset using Librispeech \cite{DBLP:conf/icassp/PanayotovCPK15} and DNS challenge dataset. All audios were sampled in 16 kHz. The learning target was set to clean and reverberant clean when the input was noisy and reverberant noisy, respectively.

The simulated dataset used the generation pipeline in \cite{chen23t_interspeech}. The only difference was that we did not include echos in this paper.

The proposed method was further evaluated on DNS challenge dataset. The training dataset was simulated on the fly following our previous configuration \cite{yu23b_interspeech}. The evaluation set is the DNS-2020 no-blind set, which contains \textit{no\_reverb} and \textit{with\_reverb} parts. The DNS-2020 no-blind set has the reference clean signal, which means that we could evaluate widely used objective metrics.

\subsection{Metrics}
Performance was evaluated by scale-invariant signal-to-noise ratio (SI-SNR)  \cite{Luo2019ConvTasNetSI}, narrow-band PESQ (PESQ-NB), PESQ-WB \cite{DBLP:conf/icassp/RixBHH01} and short-time objective intelligibility (STOI) \cite{DBLP:conf/icassp/TaalHHJ10}. PESQ-NB, PESQ-WB and STOI are computed by open-source codes \footnote{https://github.com/ludlows/PESQ}\footnote{https://github.com/mpariente/pystoi}. The full results can be found on our demo page. Here we only list PESQ-WB and SI-SNR.

\subsection{Model setup}

The STFT and iSTFT used a window size of 20 ms and a stride size of 10 ms. The detailed configuration and the corresponding numbers of model size and computational cost are listed in Table \ref{tab:hyper}. We scaled the MACs of the causal MPT network from 50M/s to 15G/s and also tested a noncausal MPT model with MACs around 23G/s. Models used gated recurrent units (GRUs) and long short-term memory networks (LSTMs) as RNN units when the MACs were below and above 1G/s, respectively.

\begin{table}[t]
    \centering
    \caption{The details of hyperparameters.}
    \scalebox{0.85}{
    \begin{tabular}{cc|c|cc|cccc}
    \toprule
    MACs/s & \# Params. & Causal & MPT & RNN  & K & B & E & C    \\
    \midrule
    50M  & 287K & \cmark   & 2+3 & GRU & 28 & 2 & 16 & 1 \\
    102M & 385K & \cmark  & 2+3 & GRU & 30 & 2 & 28 & 1 \\
    195M & 492K & \cmark  & 2+3 & GRU & 31 & 2 & 42 & 1 \\
    301M & 447K & \cmark  & 1+2+3 & GRU & 31 & 2 & 24 & 2 \\
    502M & 545K & \cmark  & 1+2+3 & GRU & 31 & 2 & 32 & 2 \\
    1.0G & 931K & \cmark  & 1+2+3 & GRU & 30 & 4 & 36 & 2 \\
    4.1G & 4.4M & \cmark  & 1+2+3 & LSTM & 30 & 5 & 56 & 2 \\
    14.0G& 12.0M & \cmark & 1+2+3 & LSTM & 30 & 6 & 96 & 2 \\
    23.2G& 14.2M & \xmark & 1+2+3 & LSTM & 30 & 6 & 96 & 2 \\
    \bottomrule
    \end{tabular}
    }
    \label{tab:hyper}
\end{table}

Models were trained using Adam optimizer on an 8 GPUs. Each model was optimized with 100K and 600K iterations on the simulated dataset and DNS challenge datasets, respectively. The checkpoint is chosen according to the performance on the validation set.

\section{Results and discussion}
\label{sec:results}


Table \ref{tab:dns} exhibits the results on the DNS challenge dataset. We compared the MPT network with the classic GRU and DeepfilterNet (DFNet) on the computational costs lower than 1G MACs/s. The original DFNet has MACs around 300M/s and was scaled to lower complexities. Meanwhile, the MPT was scaled to high computational cost ($>10$G MACs/s) by enlarging hyperparameters $E$ and $B$ and replacing GRU with LSTM. We found that the MPT network achieved competitive performance at both low and high computational costs. For example, we achieved similar results compared with state-of-art results of BSRNN \cite{yu23b_interspeech}. Due to the page limitation, we did not include all metrics for the DNS-2020 \textit{no\_reverb} / \textit{with\_reverb} test sets. The full results of Table \ref{tab:dns} can be found on the demo page.

\begin{table}[!ht]
    \centering
    \caption{Comparision of models with various computation complexity on DNS-2020 \textit{no\_reverb} / \textit{with\_reverb} test sets.}
    \scalebox{0.95}{
    \begin{tabular}{lcc|cc}
    \toprule
    Model  & Causal & MACs & PESQ-WB    & SI-SNR          \\
    \midrule
    Noisy   & - & -            & 1.82/1.58   & 9.03/0.07\\
    \hline
    GRU    & \cmark       & 101M/s & 2.50/2.58 & 16.2/15.1 \\
    DFNet \cite{DBLP:conf/icassp/SchroterERM22}  & \cmark   & 66M/s & 2.32/2.37 & 16.2/14.4 \\
    MPT    & \cmark    & 102M/s & \textbf{2.60/2.73} & \textbf{16.9/15.6} \\
    \hline
    GRU    & \cmark   & 301M/s & 2.62/2.70 & 16.3/15.2 \\
    DFNet \cite{DBLP:conf/icassp/SchroterERM22}  & \cmark   & 318M/s & 2.59/2.70 & 17.4/15.6 \\
    MPT    & \cmark   & 301M/s & \textbf{2.78/2.91} & \textbf{17.5/16.2} \\
    \hline
    GRU    & \cmark      & 1.03G/s & 2.66/2.75 & 17.0/15.6 \\
    MPT    & \cmark      & 1.05G/s & \textbf{2.97/3.18} & \textbf{18.2/16.9} \\
    \hline
    BSRNN \cite{yu23b_interspeech} & \cmark & 14.7G/s & \textbf{3.32}/3.37 & \textbf{20.5/17.9} \\
    MPT & \cmark &  14.0G/s & 3.24/\textbf{3.40} & 19.9/17.8 \\
    \hline
    BSRNN \cite{yu23b_interspeech} & \xmark & 23.4G/s & \textbf{3.45}/3.68 & \textbf{21.4/19.2} \\
    MPT    & \xmark    & 23.2G/s & 3.43/\textbf{3.69} & 21.3/18.9 \\
    \bottomrule
    \end{tabular}
    }
    \label{tab:dns}
\end{table}

\begin{table}[!ht]
    \centering
    \caption{An ablation study of MPT architecture using different computational complexity on the simulated dataset. The numbers in MPT column represent the transformer used. For example, ``1+2'' indicates that the first and the second transformer are used in each MPT block. The hyperparameters are tuned to keep the MACs around the target one.}
    \scalebox{0.9}{
    \begin{tabular}{c|ccc|c|cc}
    \toprule
    ID & MPT  & Mask & DF & MACs & PESQ-WB & SI-SNR   \\
    \midrule
    1 &1+2     & Real    & \xmark & 105M/s & 2.02 & 12.6 \\
    2 &1+2+3   & Real    & \xmark & 104M/s & 2.12 & 13.1 \\
    3 &2+3     & Real    & \xmark & 100M/s & 2.13 & 13.1 \\
    4 &2+3     & Complex & \xmark & 110M/s & 2.06 & 12.9 \\
    5 &2+3     & Real & \cmark & 105M/s & \textbf{2.21} & \textbf{13.6} \\
    \hline
    6 &1+2     & Real    & \xmark & 305M/s & 2.18 & 13.4 \\
    7 &1+2+3   & Real    & \xmark & 303M/s & 2.25 & 13.6 \\
    8 &1+2+3   & Complex & \xmark & 310M/s & 2.23 & 13.6 \\
    9 &1+2+3   & Real    & \cmark & 304M/s & \textbf{2.29} & \textbf{14.0} \\
    \hline
    10 & 1+2     & Real    & \xmark & 0.99G/s & 2.35 & 14.1 \\
    11 &1+2+3   & Real    & \xmark & 1.03G/s & 2.36 & 14.2 \\
    12 &1+2+3   & Complex & \xmark & 1.04G/s & 2.42 & 14.4 \\
    13 &1+2+3   & Complex & \cmark & 1.06G/s & \textbf{2.46} & \textbf{14.8} \\
    \bottomrule
    \end{tabular}
    }
    \label{tab:ablation}
\end{table}

We further extended the computational cost to cover the range from 50M/s to 15G/s (Figure \ref{fig:results}). We mainly tuned the hyperparameters $E$ and $B$. In detail, a model with 100M MACs/s was extended to 50M and 200M MACs/s. A model with 300M MACs/s was extended to 500M MACs/s. The model with 1G MACs/s was extended to 4G/s and 14G/s. The PESQ was the averaged score on the DNS-2020 \textit{no\_reverb} / \textit{with\_reverb} test sets. The formula of the fitting lines of MACs/s, PESQ and SI-SNR were $\text{PESQ}=0.092\times\text{log}_2\text{(MACs/s)}+2.077$ and $\text{SI-SNR(dB)}=0.36\times\text{log}_2\text{(MACs/s)}+13.87$, indicating that a half reduction of computational cost leads to a PESQ and SI-SNR degradation of around 0.09 and 0.36 on the DNS-2020 \textit{no\_reverb} / \textit{with\_reverb} test sets.

Table \ref{tab:ablation} presented the ablation results on the simulated noisy dataset. The MPT distinguished itself from other models by the MPT module and the dual-stream filtering module. Thus, the ablation study was conducted on the transformer numbers used in MPT, the mask type in the dual-stream filtering module, and whether to use deep filtering. We tested the model on three representative numbers of MACs (e.g., 100M/s, 300M/s and 1G/s) to make the conclusions more convincing. The hyperparameters $E$ and $K$ were tuned to hold the computational cost around the target complexity.

The results in Table \ref{tab:ablation} suggest three empirical discoveries. First, MPT (1+2+3) gave higher PESQ scores compared with MPT (1+2), especially on the low computational cost (ID1 vs. ID2, ID6 vs. ID7). For ultra-low complexity, using MPT (2+3) achieved better performance (ID1 vs. ID2). Second, a complex mask in the dual-stream filtering module promoted the performance when the MACs were around 1G/s (ID11 vs. ID12) and degraded the PESQ scores when the MACs were around 100M/s and 300M/s (ID3 vs. ID4, ID7 vs. ID8). Third, adding the deep filtering stream improved the performance on all computational costs.

\section{Conclusion}
\label{sec:conclusion}

This paper investigates the problem of scaling denoising models from low to high complexity. We propose a set of MPT models to cover the computational cost from 50M MACs/s to 25G MACs/s. The proposed MPT block and dual-stream filtering modules are evaluated to be effective in both low and high complexity. Meanwhile, we formulate the neural scaling laws for the first time to provide the intuitions of the performance changing following the computation cost. In future work, we intend to perform variable computational costs into one unified model and extend our model on both denoising and echo cancellation tasks.

\bibliographystyle{IEEEbib}
\bibliography{refs}

\end{document}